\documentclass[paper, final]{JFM-FLM_Au}

\lefttitle{M. Boukor, P. Tallón Marrón, R. P. T. Nguyen, J. Vétel, É. Laurendeau \& F. P. Gosselin}

\righttitle{Journal of Fluid Mechanics}

\title{Wake-Induced Drag and Phase-Reconstructed Dynamics of a Flexible Plate in Normal Flow}

\author{Maryam Boukor\aff{1}, Pedro Tall\'on Marr\'on\aff{1}, Richard Phat The Nguyen\aff{1}, J\'er\^ome V\'etel\aff{1}, \'Eric Laurendeau\aff{1} and Fr\'ed\'erick P. Gosselin\aff{1}}

\affiliation{\aff{1}Department of Mechanical Engineering, Polytechnique Montréal, Montréal, Qc, Canada}

\corresau{Fr\'ed\'erick P. Gosselin, \email{frederick.gosselin@polymtl.ca}}
\usepackage{siunitx}

\begin{document}
\maketitle

\begin{abstract}
Flexible structures in an incoming perpendicular flow typically undergo elastic reconfiguration that reduces drag; however, at higher velocities, they are prone to dynamical instabilities that entail complex wake dynamics and fluctuating loads. In this study, we investigate the wake of a thin, flexible plate clamped at its midpoint and oriented normal to an airflow, modelling reconfigurable natural systems such as trees and sea-grasses. By combining Proper Orthogonal Decomposition, Robust Principal Component Analysis, and Singular Value Decomposition with non-time-resolved Particle Image Velocimetry, we reconstruct the periodic coherent flow structures across both static and vibrating regimes. We demonstrate that structural oscillation symmetry directly dictates the wake topology. The symmetric vibration regime is characterised by two parallel 2S-type vortex shedding patterns on either side of the plate—herein termed the S-2S mode—whereas the antisymmetric regime exhibits a classic 2P-type shedding pattern. Furthermore, an impulse-based force analysis links these wake circulations to drag, revealing an additional mean drag penalty in the antisymmetric regime. Our approach offers a practical framework to extract and interpret coherent wake structures from limited temporal data, enhancing our understanding of fluid–structure interactions and informing aerodynamic load predictions.
\end{abstract}

\section{Introduction}
\label{sec:intro}

Fluid-structure interaction (FSI) exists in many forms in both natural systems and man-made structures.
In these systems, flexible structures interact with the surrounding flow to achieve functions such as propulsion or drag reduction. Examples of natural systems range from the reconfiguration of leaves in trees to reduce drag in strong winds, where porosity and structural flexibility play a role in how the structure interacts with the flow \citep{vogel1989drag,gosselin2010drag,baskaran2023reconfiguring,gehrke2025coupled}, to the bending of fins and wings that enables efficient propulsion in fish and birds \citep{taylor2003flying, gehrke2025highly}, and even to the jet propulsion of jellyfish, which swim by rhythmically contracting and expanding their bell \citep{costello2008medusan,dabiri2005flow}.
However, when flexibility or flow speed increases beyond certain limits, the beneficial effects of deformation such as drag reduction and propulsion enhancement can break down \citep{heathcote2008effect,eloy2007flutter,shelley2011flapping}.

To quantify the efficacy of this drag reduction, \cite{vogel1989drag} introduced an empirical measure known as Vogel's exponent ($\mathcal{V}$), which characterises how the aerodynamic drag of a flexible body scales with flow velocity as $U^{2+\mathcal{V}}$, increasing more slowly than the classical square of the incoming velocity ($U^2$). Underlying this macroscopic drag reduction is a fundamental scaling of the body's elastic deformation. Analytical and experimental works by \cite{alben2002drag} and \cite{gosselin2010drag} established that this structural reconfiguration and its resulting drag scaling are governed by a dimensionless Cauchy number, reflecting the balance between hydrodynamic pressure and structural bending stiffness. 
In the limit of large deformations, the structure adopts a universal self-similar shape whereby both its projected frontal area and the size of its wake's recirculation zone scale dynamically with the Cauchy number.

To investigate these mechanisms in a controlled manner, many studies have turned to simplified laboratory configurations in which the body geometry and its orientation relative to the incoming flow can be precisely prescribed \citep{alben2002drag,schouveiler2006rolling,gosselin2010drag,baskaran2023reconfiguring}.

A useful way to categorise FSI systems is by their alignment with the flow. In cross-flow configurations, the structure is oriented perpendicular to the flow, behaving like a bluff body and promoting large-scale flow separation and a pronounced wake; in these cases, fluid loading is typically dominated by pressure drag \citep{roshko1955wake,blevinsFlowinducedVibration1990}. In contrast, axial-flow configurations involve structures aligned with the flow direction, resembling streamlined bodies where the flow remains largely attached and drag is primarily due to skin friction, with pressure effects playing a smaller role \citep{lyon1934drag,paidoussisFluidstructureInteractionsSlender1998a}.

Several studies have analysed the flow field in cross-flow configurations, particularly focusing on wake dynamics and vortex shedding. The type of wake observed behind a body depends strongly on its geometry, boundary conditions, and interaction with the flow \citep{wiliamson1996vortex}. For rigid bluff bodies such as circular cylinders, the wake typically transitions from a steady symmetric flow at low Reynolds numbers to a periodic von Kármán vortex street as the Reynolds number increases. This classical behaviour has been extensively documented, notably by \cite{williamson1988vortex}, who classified wake patterns based on Reynolds number and body motion. In this classification, 'S' denotes the shedding of single vortices per cycle, while 'P' indicates vortex pairs. For a spring-mounted cylinder free to oscillate in the transverse direction, as the flow speed increases and the vortex-shedding frequency can synchronise with the oscillations, leading to a lock-in regime. In this lock-in range, large vibration amplitudes are expected alongside a transition between distinct wake modes, such as 2S, 2P, and P+S. 
For a cylinder vibrating in the streamwise direction, different vortex patterns can be expected \citep{ongoren1988flow, jauvtisVortexinducedVibrationCylinder2003, cagneyWakeModesCylinder2013, cagney2014streamwise}. Notably, an S-I vortex pattern can be encountered where a pair of vortices of opposite signs are shed simultaneously from both sides of the cylinder for every cycle. Different wake structures are associated with different flow-induced instabilities.

By contrast, axial-flow configurations, particularly for slender and flexible structures, are commonly approached from a fluid–elastic stability perspective, where flutter or divergence emerges from distributed loading along the structure. A canonical example is the work of \cite{zhang2000flexible}, who investigated an upstream-clamped flexible filament in a two-dimensional flowing soap film and demonstrated the existence of two stable states: a stretched-straight configuration aligned with the flow and a self-excited large-amplitude flapping state beyond a critical velocity. Subsequent studies have shown that streamwise-aligned flexible structures primarily lose stability through flutter, with critical thresholds set by the balance between aerodynamic loading, structural stiffness, and inertia \citep{tang2003flutter,souilliez2006experimental,eloy2007flutter,eloy2008aeroelastic}. Beyond the onset of instability, the wake response can be characterised across different flapping and oscillatory regimes. \cite{schnipper2009vortex} mapped a rich phase diagram of vortex wake topologies behind a flapping foil in a soap film, identifying patterns ranging from classic 2S and 2P shedding to more complex configurations involving up to 16 vortices per oscillation cycle.

Between these two canonical configurations—cross-flow and axial-flow—there exists an important class of hybrid fluid–structure interaction systems, which exhibit traits of both. 
For example, the inverted flag transitions from a straight state pointing against the flow---an axial flow problem---to a strongly deflected mean or bluff state and then to self-excited flapping as flow speed increases, accompanied by marked changes in wake topology \citep{kim2013flapping,kim2017dynamics,tavallaeinejad2020instability,patel2025fluid,giri2025dynamics,naik2026energy, zhangEnergyConversionDynamics2026}. 
Then there are flexible structures that progressively reconfigure and transition from cross-flow-dominated states with large mean deformation to streamwise-aligned configurations that exhibit self-excited vibrations, where flow-induced deformation continuously alters the structure's effective orientation and geometry. 
\cite{leclercqDoesFlutterPrevent2018} studied analytically a cantilevered flexible plate in a uniform flow that is initially perpendicular to the flow, but which reconfigures and aligns with the flow as the flow speed increases. By performing a stability analysis about the static reconfigured shapes, they demonstrated that while static reconfiguration is highly beneficial to reduce aerodynamic drag, the onset of flutter instability inherently limits this drag reduction. This results in a coexistence and interplay between bluff-body wake features and axial-flow flutter mechanisms. 

\cite{boukor2024flutter} investigated experimentally the balance between drag reduction and dynamic instability in a flexible plate exposed to normal flow.  Their wind tunnel experiments and force measurements showed that maximal drag reduction coincides with the onset of a flutter instability, characterised by transitions from static to symmetric and then to antisymmetric vibration modes. 
While this study clarified the conditions for instability onset and modal transitions, it did not provide direct insight into the wake dynamics associated with each vibration regime. In particular, the connection between structural oscillations and vortex shedding mechanisms remains poorly understood.

Several approaches have been developed to reconstruct and analyse unsteady wake dynamics from limited measurements. One well-known strategy is the wake-capture approach, which seeks to model the influence of past wake structures on the current state of the system, often requiring time-resolved data to resolve vortex trajectories and memory effects \citep{nabawy2023simple,birch2003influence}. However, these methods are inapplicable to non-time-resolved PIV (Particle Image Velocimetry) data. In such cases, data-driven decomposition techniques like Proper Orthogonal Decomposition (POD) offer an alternative framework to reconstruct coherent flow structures and assess their spatial organisation \citep{lumley1967structure}. By leveraging the dominant energetic modes extracted from a set of snapshots, POD enables the reconstruction of the flow field and the identification of modal signatures linked to specific dynamical regimes \citep{lumley1967structure,taira2017modal,fernando2014modal}. This technique has been adopted in experimental studies to interpret wake dynamics in systems where temporal resolution is limited \citep{kourentis2012uncovering}. While it provides a compact representation of energetic flow structures, its modes may combine contributions from the mean flow, coherent vortical structures and measurement noise, which can complicate physical interpretation. To further isolate dynamically relevant features of the wake, Robust Principal Component Analysis (RPCA) has emerged as a complementary tool for flow decomposition. RPCA separates the data into a low-rank component, which isolates the coherent large-scale motions, and a sparse component, which accounts for the undesirable effects of outliers and measurement noise, as well as stochastic turbulence contributions. This separation has been shown to improve the identification of coherent wake features in experimental and numerical studies of unsteady flows \citep{scherl2020robust}.

While previous studies, such as \cite{boukor2024flutter}, have identified the structural boundaries of flutter instability for a flexible plate in normal flow, the underlying unsteady flow physics—specifically the vortex shedding mechanisms and their direct link to aerodynamic forces—remain unresolved.
To address this gap, the present study comprehensively characterises the fluid-structure coupling and wake dynamics of a flexible normal plate. The core contributions of this work are threefold:
\begin{itemize}
    \item We achieve a full reconstruction of the periodic wake dynamics from non-time-resolved PIV measurements across four distinct regimes: static reconfiguration, symmetric vibration, an intermediate transition regime, and anti-symmetric vibration.
    \item We show that the scaling laws governing drag and shape of a flexible plate normal to the flow also applies to the size of the wake recirculation zone. Moreover, we show how the emergence of dynamical instabilities either preserves or disrupts the time-averaged drag scaling.
    \item We explicitly link each structural vibration regime to its corresponding vortex shedding topology, revealing how symmetric and anti-symmetric structural motions dictate fundamentally different wake patterns.
\end{itemize}

The remainder of this paper is organised as follows. Section 2 details the experimental setup and the flow reconstruction methodology. Section 3 presents the results, including the wake flow characterization, modal decomposition, periodic vorticity patterns, and the impulse-based drag analysis. Finally, concluding remarks are provided in Section 4.

\section{Methodology}
\subsection{Experimental Setup}
\begin{figure}
\centering
\normalfont
\def\svgwidth{0.8\textwidth}
\input{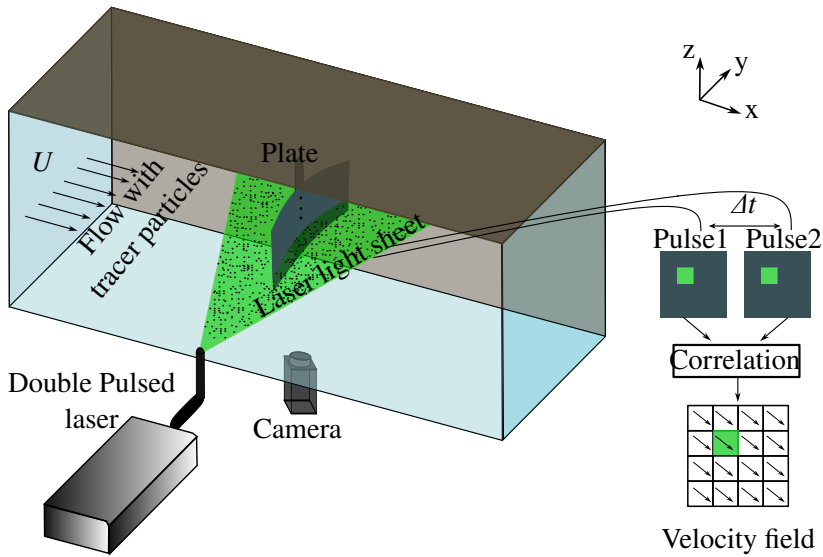}
\caption{Schematic of the experimental PIV setup. The specimen is positioned in the wind tunnel test section and illuminated by a laser sheet defining the measurement plane. Particle images recorded by the camera are processed using cross-correlation to obtain velocity fields. }
\label{fig:setup}
\end{figure}

The force and PIV experiments were conducted in a closed-circuit wind tunnel (Model 407B, Engineering Laboratory Design Inc.) at the Department of Mechanical Engineering, Polytechnique Montréal, as illustrated in Figure \ref{fig:setup}. The wind tunnel features a working section measuring \SI{121.9}{\centi\meter} (length) $\times$ \SI{61.0}{\centi\meter}  (width) $\times$ \SI{61.0}{\centi\meter} (height), with a \SI{150}{HP} motor capable of producing airspeeds ranging from 3.0 \si{\meter\per\second}  to 91.4~ \si{\meter\per\second}. The flow speeds investigated in this study correspond to length-based Reynolds numbers $Re_L$ ranging up to approximately $7.5 \times 10^4$. The test section is equipped with plexiglass windows on the bottom and sides for optical flow visualization and a wooden panel on the top, which supports a load cell. The six-axis load cell (Gamma, ATI Technologies) measures drag and lift forces.

Specimens of flexible rectangular flat plates were tested by fixing them individually at their midline using steel wires to an aluminum mast attached to the force balance. The specimens were fabricated from PTFE sheets, cut to specific dimensions using a CO$_2$ laser cutter.
The specimen used in this study is a flexible rectangular plate, with a length $L$ = 80~\si{\milli\meter}, width $w$ = 40~\si{\milli\meter}, and thickness $h$ = 0.076~\si{\milli\meter}. Its flexural rigidity is $D$ = 0.18~\si{\newton\milli\meter}, area density $m_s$=0.108~\si{\kilogram\per\meter\squared}
and the dimensionless mass number $\rho_f L/m_s=0.88$, with $\rho_f$ denoting the fluid density. The specimen dimensions were selected to match the available field of view of the PIV setup, ensuring that the entire plate and its near wake could be captured simultaneously. In addition, the structural parameters were chosen to allow the exploration of all expected dynamical regimes, ranging from static reconfiguration to symmetric and antisymmetric vibration.

Flow visualization was performed using a PIV system (nanopiv, LaVision) equipped with an Nd:YAG pulsed laser (Model NANO L 200-15, Litron Lasers) operating at 532~\si{\nano\meter} with a maximum repetition rate of \SI{15}{\hertz}.  The laser beam, with a diameter of $1/4~\text{inch}$, generated a light sheet. The wind tunnel airflow was seeded with Di-Ethyl-Hexyl-Sebacate (DEHS) aerosol particles, atomized and dispersed uniformly for \SI{20}{\minute} at 2~\si{\meter\per\second}. The particles, with an average diameter of \SI{0.3}{\micro\meter}, acted as tracers for the flow.
Tracer particle motion was captured using a LaVision Imager sCMOS camera mounted on an $XZ$ translation stage beneath the test section. The camera operated in double-frame, double-exposure mode, with a resolution of $2560 \times 2160$ pixels. A Sigma 150\si{\milli\meter} f/2.8 EX DG OS HSM Macro lens was used for imaging, while the laser was positioned on a wheeled optical table to align the light sheet perpendicular to the specimen’s midline. PIV experiments and post-processing were performed using DAVIS 8.2.3 software. The laser operated at 100\% power for both pulses, with the pulse interval optimized to $dt = 60\,\mu\mathrm{s}$
using the software's \textit{dt optimizer} tool to balance particle density and spatial resolution. Velocity fields were obtained over a \SI{144}{\milli\meter}  (length) $\times$ \SI{120}{\milli\meter} (width) field of view. Multi-pass cross-correlation was applied, with interrogation windows reduced from $64 \times 64$ pixels to $32 \times 32$ pixels, yielding $21,600$ vectors ($N_x = 160$, $N_y = 135$). A peak-to-peak ratio (PPR) technique was used to eliminate vectors with high uncertainty. Batch processing allowed the efficient handling of $1000$ PIV images per configuration, with full processing taking up to $8~ \mathrm{hours}$. To minimise reflections, the specimens were painted with a black marker, and matte black cardboard and adhesive tape were applied to the upper and side walls of the wind tunnel. These measures reduced light scattering without altering the specimens’ weight or rigidity. Even with these measures, residual artefacts persisted in the PIV fields, arising from out-of-plane particle motion, laser-sheet non-uniformities, and reflections that could not be fully eliminated. A dedicated post-processing approach was therefore implemented to mitigate these effects and obtain coherent flow structures, as discussed in the next section.
The camera’s limited field of view restricted the observation of downstream flow structures to $x/L \leq 3.5$, where $L$ is the plate length (with $L$ = 80\si{\milli\meter}).

\subsection{Flow reconstruction}
\label{sec:method}
\begin{figure}
\normalfont
\def\svgwidth{1\textwidth}
\input{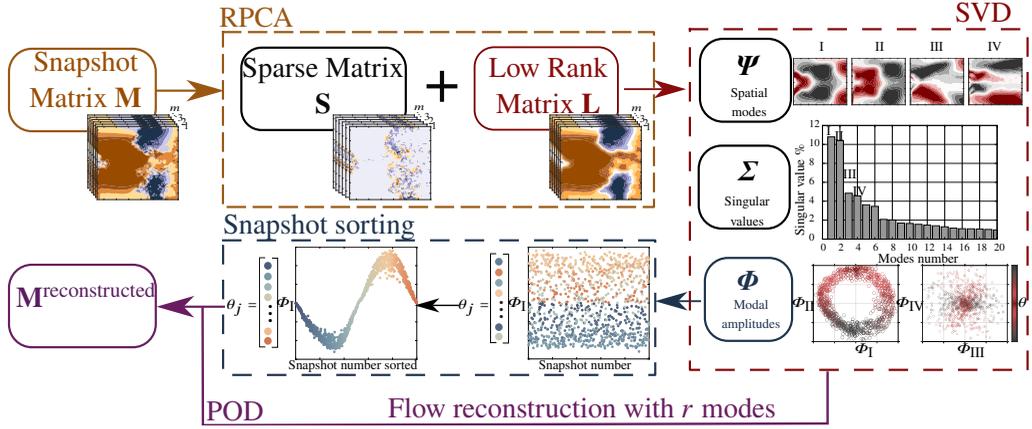}
\caption{Schematic of the modal decomposition methodology. The procedure involves RPCA for separating coherent dynamics from noise, SVD for extracting dominant modes, and POD with snapshot sorting to organize snapshots in phase. The reconstruction step yields mode shapes, temporal coefficients, and phase-plane portraits that characterize the flow dynamics.}
\label{fig:methodology_schematic}
\end{figure}

The methodology adopted to postprocess the non-time resolved PIV data involves different steps to extract, clean, filter the data and reconstruct the flow as schematised in Fig. \ref{fig:methodology_schematic}. This procedure includes the application of RPCA \citep{candes2011robust}, SVD and POD, followed by a flow reconstruction that considers the sorting of the angle between the principal modes. First, preprocessing of the vector field from PIV data is done by removing the part of the plate specimen from the snapshots. 
For snapshot $j$, the corresponding column vector $\mathbf{a}_j$ is built by stacking the $N$ streamwise velocity component $u$ (in the $x$-direction) and $N$ transverse velocity component $v$ (in the $y$-direction), where $N = N_x \times N_y$ is the number of spatial grid points in each 2D field and each grid point is associated with an $(x,y)$ coordinate:
\begin{equation}
    \mathbf{a}_j = 
    \begin{bmatrix}
        u_{1j} \\
        \vdots \\
        u_{Nj} \\
        v_{1j} \\
        \vdots \\
        v_{Nj}
    \end{bmatrix},
    \qquad j = 1,\dots,m.
\end{equation}
The data matrix $\mathbf{M}$ is then defined by combining the velocity snapshots as
\begin{equation}
    \mathbf{M} = \bigl[\, \mathbf{a}_1 \;\; \mathbf{a}_2 \;\; \cdots \;\; \mathbf{a}_m \,\bigr].
\end{equation}
Here, $m=1000$ is the number of snapshots.

To separate the coherent structures from noise and outliers, RPCA was applied to data matrix $\mathbf{M}$. This method decomposes the matrix $\mathbf{M}$ into two components:
\begin{equation}
   M_{ij} = L_{ij} + S_{ij},  
\end{equation}
where $i = 1,\dots,2N$ denotes the number of degrees of freedom corresponding to the concatenated velocity components. $\mathbf{L}$ is the low-rank matrix, representing coherent and dominant flow structures, and $\mathbf{S}$ is the sparse matrix, capturing localized perturbations and experimental noise.

SVD is applied to extract the dominant spatial and temporal modes:
\begin{equation}
    L_{ij} = \sum_{k=1}^{r}\Psi_{ik}\,\sigma_k\,\Phi_{jk},  
    \label{eq: matrix L}
\end{equation}
where $r$ indexes the retained SVD modes. The matrix $\boldsymbol{\Psi}$ contains the spatial modes, the diagonal matrix $\boldsymbol{\Sigma}$ contains the singular values $\sigma_k$ associated with the energy of each mode, and the matrix $\boldsymbol{\Phi}$  describes the temporal coefficients corresponding to the modal amplitudes.

To reconstruct the flow's periodic behaviour, the first two temporal modes  $\Phi_I$ and $\Phi_{II}$ were used to calculate the phase angle $\theta_j$ for each snapshot: 
\begin{equation}
\theta_j = \arctan\left(\frac{\Phi_{j2}}{\Phi_{j1}}\right).
\end{equation}
This phase extraction inherently assumes that the first two POD modes act as a quasi-harmonic convective pair, forming a circular limit-cycle in phase space; this assumption is statistically validated using a Randomized Dependence Coefficient (RDC) analysis detailed in Appendix \ref{sec:appendix2}. This angle characterises the position of each snapshot in a reduced phase space. Sorting the phase angles $\theta_j$ in ascending order defines a permutation $\pi$ of the snapshot index such that $\theta_{\pi(1)} \le \theta_{\pi(2)} \le \dots \le \theta_{\pi(m)}$.
Applying this permutation organises the snapshots into a phase-ordered sequence corresponding to a single oscillation cycle. The same permutation is applied to the temporal coefficients, yielding a reordered temporal coefficient matrix $\boldsymbol{\hat\Phi}$, defined as:
\begin{equation}
\hat{\boldsymbol{\Phi}}_{jk} = \boldsymbol{\Phi}_{\pi(j)k}.
\end{equation}
The flow field is reconstructed using the first $r$ most energetic POD modes by combining the spatial modes with the phase-ordered temporal coefficients, yielding a phase-consistent representation of the flow:
\begin{equation}
M^{\mathrm{reconstructed}}_{ij}
= \sum_{k=1}^{r} \Psi_{ik}\,\sigma_k\,\hat{\Phi}_{jk}.
\label{eq: reconstructed M}
\end{equation}
This reconstructed matrix retains the dominant spatial structures while enforcing temporal coherence, enabling clearer interpretation of the underlying flow dynamics.

In this study, the fluctuations in the flow downstream of the specimen are assumed to be predominantly periodic, meaning that the dominant flow modes exhibit oscillatory behaviour over time. However, due to the lack of temporally resolved experimental data, crucial temporal information between consecutive snapshots is lost. This limitation prevents a true reconstruction of the time evolution of coherent structures in the flow.
To address this and obtain cleaner spatial modes, a sinusoidal model was used to idealise the temporal behaviour of the flow modes. This approach assumes that the fluctuations can be approximated as sinusoidal functions of the phase $\theta \in [0,2\pi]$. By applying nonlinear regression to the phase-sorted temporal modes $\hat{\Phi}$, each mode $k$ is approximated independently by a sinusoidal function, the fitting procedure estimates three coefficients: the amplitude $a_k$, the angular frequency $b_k$, and the phase shift $c_k$.
These coefficients are then used to construct idealised temporal coefficients by evaluating the fitted sinusoid along the reordered snapshot index $j$, yielding
\begin{equation}
\hat{\Phi}^{\mathrm{ideal}}_{jk}
= a_k \sin\!\left(b_k\, j + c_k\right).
\end{equation}
The fitting process ensures that the reconstructed modes accurately capture the oscillatory behaviour of the system while maintaining consistency with the assumed periodicity.
The fitted sinusoidal function is used directly to reconstruct the velocity field, providing a smoothed and idealised representation of the flow dynamics. The procedure provides a more continuous phase representation of the oscillation and yields a clearer visualisation of the flow field. Importantly, the coherent structures and their associated physical effects are preserved, ensuring that the essential dynamics of the system are faithfully represented. This methodology therefore bridges the gap between the limited temporal resolution of the experimental dataset and the underlying periodic behaviour of the flow, enabling a more robust characterisation of its dynamics.

\subsection{Image tracking}
The structural motion of the plate was extracted from video recordings using a transformer-based model that enables accurate two-dimensional tracking of multiple points in video sequences (CoTracker, Meta AI) \citep{karaev2024cotracker}. Unlike conventional tracking approaches, it performs joint tracking by exploiting the interdependence between points, thereby improving robustness. This method enabled the reconstruction of the plate deformation over the reordered photograph snapshots.

To quantify the plate motion, the images were first sorted to ensure smooth displacement between successive frames, making the tracking more robust (Figure~\ref{fig:methodology_schematic}). The tracking algorithm then followed a set of markers defined manually along the plate’s edge in the first snapshot. From the $P$ tracked points $\{(x_l^{(j)}, y_l^{(j)})\}_{l=1}^{P}$, we compute a length-weighted centroid of the polyline for each snapshot $j$. Consecutive points define segments with midpoints $(\bar{x}_\ell, \bar{y}_\ell)$ and arc lengths $s_\ell$, so the centroid coordinates are given by:
\begin{equation}
x_0^{(j)} =
\frac{\sum_{\ell=1}^{P} s_\ell \, \bar{x}_\ell}
     {\sum_{\ell=1}^{P} s_\ell},
\qquad
y_0^{(j)} =
\frac{\sum_{\ell=1}^{P} s_\ell \, \bar{y}_\ell}
     {\sum_{\ell=1}^{P} s_\ell}.
\end{equation}
The resulting centroid trajectory is obtained as a discrete sequence $y_0(j)$, where $j=1,\dots,m$ denotes the snapshot index. Since the snapshots correspond to one ordered oscillation cycle, the index can be initially mapped to a phase angle $\theta_j = 2\pi j / m$.

To synchronize this phase index with physical time, the fundamental flapping frequency $f$ was identified from the synchronous force-balance signal via FFT. We then set $\theta(t) = 2\pi f\,t$, so that the snapshots distributed over one cycle correspond to time instances $t_j = \frac{j}{m}\,\frac{1}{f}$. This ensures strict temporal consistency between the phase-ordered PIV sequence and the physical period of the structure.
The centroid motion is then idealised as a pure cosine function of phase, equivalently written in time as:
\begin{equation}
    \label{eq:centroid}
y_0(t) \;\approx\; Y_0 \cos(2\pi f\,t),
\end{equation}
where $Y_0$ is the oscillation amplitude. Phase shifts and offsets are neglected here since the snapshots are phase-ordered and the mean displacement is negligible compared to the oscillatory component. The finalized oscillatory trajectory $y_0(t)$ is subsequently retained for the aerodynamic force analysis.

The experimental data in terms of velocity fields extracted from PIV and the Matlab code to perform the post-treatment are accessible online \citep{zotero-item-23149}.

\section{Results}
\begin{figure}
\centering
\normalfont
\def\svgwidth{1\textwidth}
\input{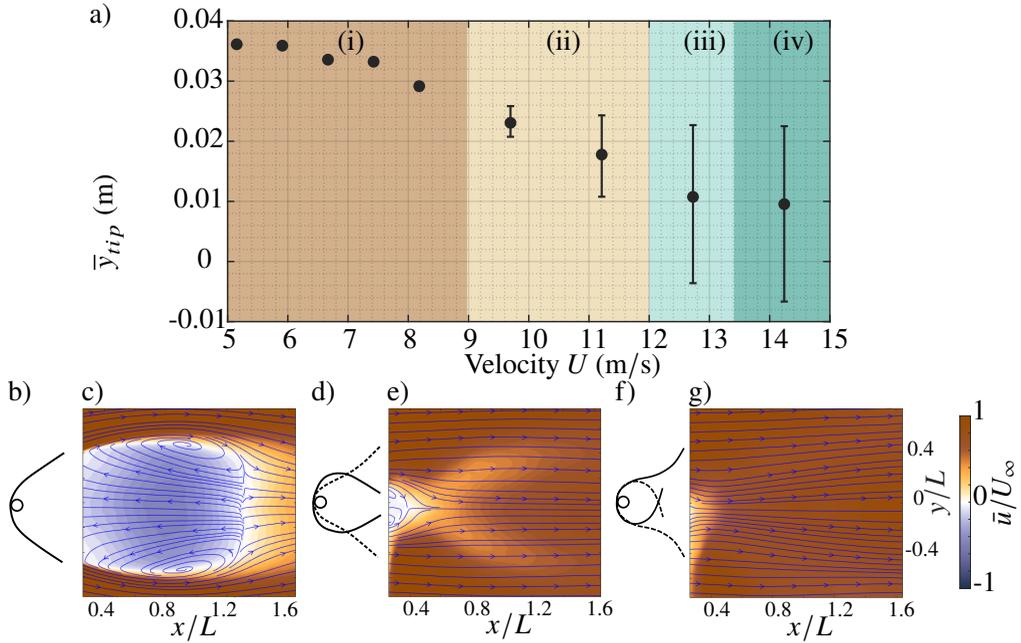}
\caption{Evolution of plate deformation and mean wake organization across flow regimes. (a) Mean $y$-position of the tip with respect to the symmetry plane $\overline{y}_{tip}$. The lower and upper limits of the tip motion are defined by the 10th and 90th percentiles. (b-g) Deformation of the flexible plate \textit{green plastic 8-4} $(L=$ \SI{8}{\centi\meter}, $w=$\SI{4}{\centi\meter}) and the corresponding mean streamwise velocity field in its wake, for different regimes: b-c) Static reconfiguration regime (i) at $U$ = 6.6~\si{\meter\per\second}, d-e) Symmetrical vibration (ii) at $U$ = 11.2~\si{\meter\per\second}, f-g) antisymmetrical vibration (iv) at $U$ = 14.2~\si{\meter\per\second}. Plate shapes are extracted from the original PIV photographs by tracing the plate contour. Velocity component $u$ is normalized by the inflow velocity $U$, and displayed using colour contours with streamlines.}
\label{fig:mean flow}
\end{figure}

We focus here on the wake dynamics associated with four flow regimes observed as the inflow velocity increases. The plate exhibits distinct structural kinematics (Figure~\ref{fig:mean flow} (b), (d) and (f)), while the corresponding mean $y$-position of the tip with respect to the symmetry plane is shown in Figure~\ref{fig:mean flow}(a). This figure summarises the evolution of the plate tip response with inflow velocity. The tip displacement along $y$ (transverse direction) is extracted from the experimental image sequence using the \textit{CoTracker} tracking method. For each velocity case, the marker indicates the time-averaged tip position, while the vertical bars represent the 10th and 90th percentiles of the tip motion over the sequence. As the inflow velocity $U$ increases, the mean tip position $\overline{y}_{tip}$ decreases gradually, reflecting the progressive static reconfiguration of the plate under increasing aerodynamic loading. In this regime, the plate deforms steadily without significant oscillations (Figure~\ref{fig:mean flow} (b)), and the spread of the tip motion remains limited. Beyond this reconfiguration dominated response, a marked increase in the spread of the tip motion is observed, indicating the onset of vibratory instabilities and the transition to the dynamic regimes. The shaded regions delineate the four flow regimes identified from the structural kinematics extracted from the experimental image data and from the drag-force changes, quantified using the coefficient of variation \citep{boukor2024flutter}: (i) static reconfiguration, (ii) symmetric vibration, (iii) transition regime, and (iv) antisymmetric vibration. 
For the case presented in this study, two distinct instability thresholds were identified. The first critical velocity, at which the onset of instability occurs, lies between $U = 8.2$ and 9.7~\si{\meter\per\second}, and corresponds to the development of a symmetric vibration mode where both sides of the plate flap in phase (Figure \ref{fig:mean flow}d). A second critical velocity is observed near $U$ = 12.7~\si{\meter\per\second}, marking the transition from symmetric to antisymmetric vibration, where the two sides of the plate oscillate out of phase (Figure~\ref{fig:mean flow}f).

\subsection{Mean wake flow characterisation}
To characterise the influence of the plate dynamics on the wake, Figure~\ref{fig:mean flow} (c), (e) and (g) presents the mean streamwise velocity field ($\Bar{u}/U$) for the three flow regimes: static reconfiguration prior to the onset of instability, symmetric vibration before the transition, and antisymmetric vibration after the transition.
These fields, obtained from PIV measurements, highlight the changes in wake structure associated with the different deformation states of the plate.

As observed in Figure \ref{fig:mean flow} (c), at low speed, the streamwise velocity field reveals a pronounced velocity deficit along the centreline. The wake is characterised by a symmetric and well-defined recirculation region immediately downstream of the plate, bounded laterally and vertically. Its structure is typical of bluff-body wakes, consistent with separated but steady flow behind a deformed yet stationary configuration. As the air speed increases (see Figure \ref{fig:mean flow}e), the recirculation region becomes smaller and more confined.
As the instability occurs and the symmetrical vibration appears, the wake remains mirror-symmetric and the mean streamwise velocity field reveals a closed teardrop-shaped recirculation bubble directly behind the plate. 
The plate oscillates symmetrically with an opening–closing kinematic pattern, which alters wake entrainment. Consequently, the mean streamwise velocity deficit in $u$ shifts away from a centreline-aligned pattern and forms a $90^\circ$-clockwise rotated V-shape.
With further increase in flow velocity, the regime transitions to an antisymmetric vibration. The plate undergoes opposite-phase bending about the centreline (Figure \ref{fig:mean flow}f), alternating between upward and downward deflected states. However, as seen in Figure \ref{fig:mean flow}g, the mean streamwise velocity field still produces a mean wake that remains symmetric, while the recirculation zone observed at lower velocities disappears.
It is worth noting that the white patch visible at the left edge of the plot, which appears non-symmetric with respect to the centreline, is not a flow feature but a visualisation artefact. It results from a void in the velocity field caused by the laser shadow projected by the upper tip of the plate. This masked region should therefore be disregarded in the interpretation of the wake structure.

As the plate reconfigures, its mean tip position moves closer to the plate centreline in the static reconfiguration regime (i), as shown in Figure \ref{fig:mean flow}(a).  \cite{alben2002drag} and \cite{gosselin2010drag} have shown that reconfigured plate profiles approach a nearly universal shape when lengths are rescaled using a Cauchy number based factor. The Cauchy number is defined as
\begin{equation}
    C_y = \frac{\rho_f U^2 L^3}{16 D},
\end{equation}
where $\rho_f$ is the air density and $D$ is the flexural rigidity. In the large deformation regime $C_y\gg1$, the plate deformation asymptotically approaches self-similar profiles and different observations collapse when spatial coordinates are stretched by $\xi=C_y^{1/3}$. This scaling originates from the balance between elastic restoring forces and fluid dynamic loading, and has been shown to effectively collapse deformed plate profiles across a wide range of $C_y$. Following their approach, we show that not only the plate shape, but also the associated flow features, can be collapsed using the same $C_y^{1/3}$ rescaling. 

\begin{figure}
\centering
\normalfont
\def\svgwidth{1\textwidth}
\input{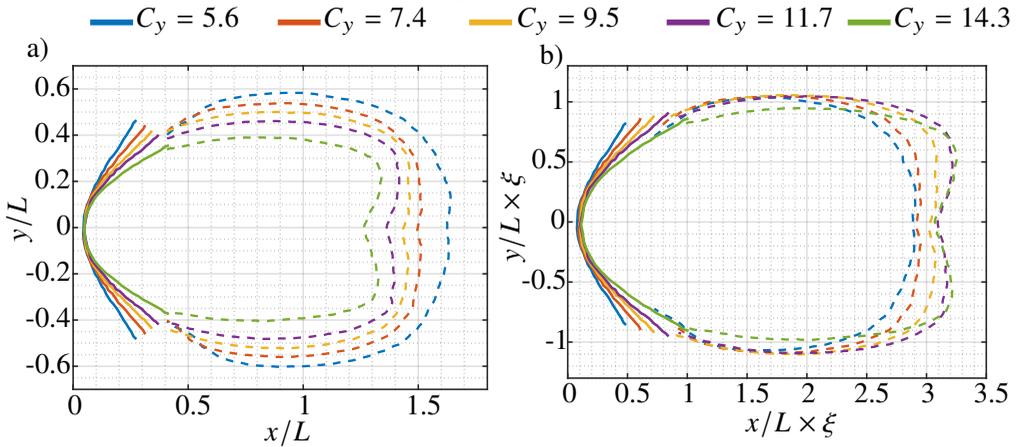}
\caption{Deformed shape of a reconfigured plate and its recirculation zone in the static regime with: a) dimensions normalized by the rigid plate length $L$, b) dimensions stretched by $\xi=C_y^{1/3}$. The recirculation zone is delimited by the time-averaged $\overline{u}=0$ contour line. }
\label{fig:scale factor}
\end{figure}

Figure \ref{fig:scale factor} presents the shapes of the reconfigured plate obtained from tracking, together with the recirculation contours ($\overline{u}=0$) extracted from the mean streamwise velocity field at different Cauchy Number values. In panel a), the lengths are normalized by the rigid plate length $L$, while in panel b) they are further stretched by the factor $\xi=C_y^{1/3}$. 
From Figure \ref{fig:scale factor}(a), it can be observed that as the flow velocity increases, both the length and the width of the recirculation zone decrease. The transverse size of the bubble varies in proportion to the reconfigured plate length. When applying the stretching factor in Figure \ref{fig:scale factor}(b), the contours of the recirculation zone ($\overline{u}=0$) collapse effectively under the same scaling, particularly at the higher Cauchy numbers achieved in this study (e.g.\ for $C_y = 9.5,\; 11.7,\; 14.3$). This indicates that, even at moderate Cauchy numbers, the wake dynamics adjust in a way that is consistent with the universal scaling. In contrast, the plate profiles collapse partially onto a single curve. This limited collapse is attributed to the relatively low values of the Cauchy number reached in these experiments ($C_y \approx 5{-}20$), for which the plate deformation remains modest. A similar trend was reported by \cite{gosselin2010drag}, who showed that at low Cauchy numbers the plate shapes do not fully collapse because the reconfiguration is not yet pronounced. These observations indicate that, while the wake rapidly conforms to the universal scaling, the structural deformation requires larger $C_y$ values to converge toward the theoretical universal shape.

\subsection{Modal decomposition of wake dynamics}
\begin{figure}
\centering
\normalfont
\def\svgwidth{1\textwidth}
\input{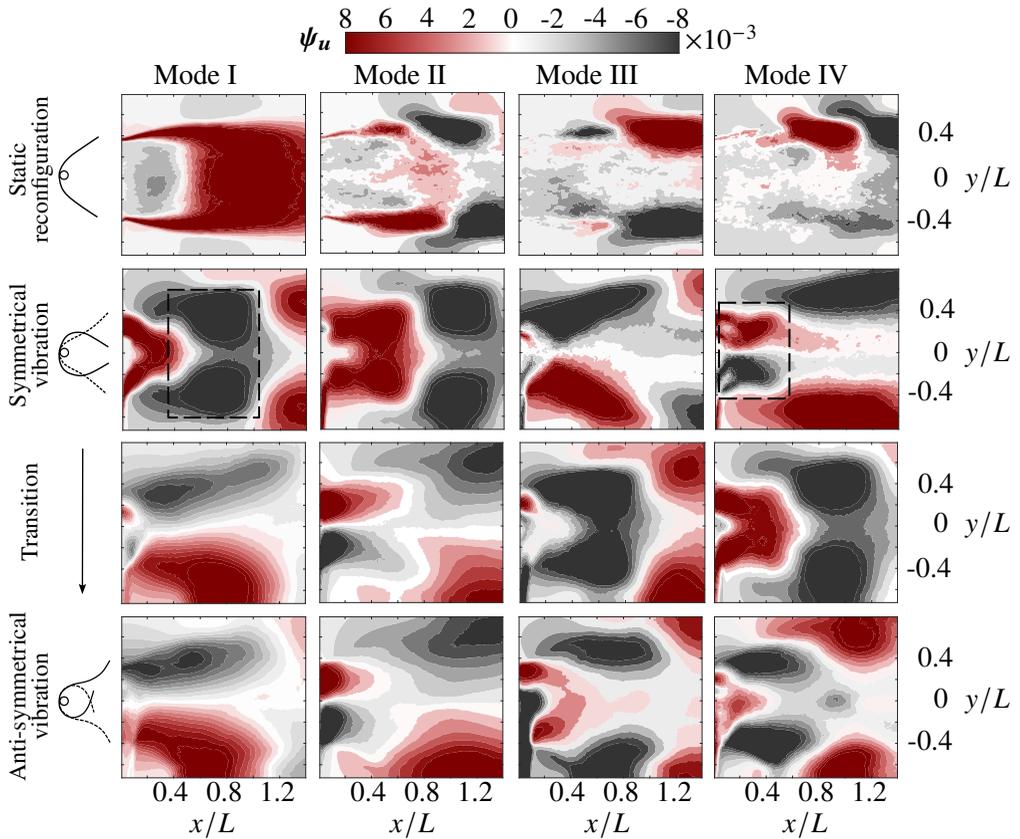}
\caption{The first four POD modes of the streamwise velocity $u$, denoted as $\boldsymbol{\psi_u}$, of the green plastic 8-4. a) The first line represents the static reconfiguration regime at $U$ =8.2~\si{\meter\per\second}. b) The second line represents the symmetrical vibration at $U$ = 11.2 ~\si{\meter\per\second}. c) The third line represents the transition from the symmetrical vibration to the antisymmetrical vibration at $U$ = 12.7~\si{\meter\per\second}. d) The fourth line represents the anti-symmetrical vibration regime at $U$ = 14.2~\si{\meter\per\second}. The dashed boxes highlight the dominant coherent structures associated with the symmetry or antisymmetry of the wake organisation.}
\label{fig:mode shapes}
\end{figure}

The analysis of mean velocity fields and recirculation bubble scaling highlights the overall organisation of the wake across regimes. To further capture the unsteady dynamics associated with vortex shedding and plate vibrations, we next apply POD to PIV data as detailed in section \ref{sec:method}. 
Figure \ref{fig:mode shapes} shows the first four mode shapes of the streamwise velocity obtained for each of the observed four regimes. These mode shapes provide insights into the evolution of coherent structures in the flow and their interaction with the deformations of the flexible plates. The POD results of the streamwise velocity field reveal two distinct types of spatial structures, defined relative to the horizontal symmetry axis ($x$-axis): a symmetric structure and an antisymmetric one. In the symmetric structure case, the velocity profiles and their associated energy distributions are mirrored and identical across the symmetry axis. In contrast, antisymmetric structure modes exhibit similar velocity profiles on either side of the axis, but with opposite signs, indicating an antisymmetric spatial structure in the wake.

In the static reconfiguration regime, Mode I represents the dominant large-scale wake fluctuation, mainly a modulation of the velocity deficit behind the plate. This mode could possibly be related to vortices shed with $y$-direction vorticity. Mode II is predominantly symmetric and is concentrated near the wake edges, indicating coherent fluctuations associated with the shear-layer region. Modes III and IV contain smaller, localized patches concentrated near the wake edges, indicating secondary wake fluctuations developing within the shear layer region.
In the symmetrical vibration regime, the first two modes remain symmetrical, reflecting synchronized flow-structure coupling on both sides of the plate, while the third and fourth shift to antisymmetrical, indicating alternating energy contributions. 
During the transition regime, a marked shift occurs. The first two modes become antisymmetric, indicating an alternating organisation of the wake fluctuations. The third and fourth modes return to symmetric structures similar to those observed in the symmetric vibration regime, showing that symmetric components remain present but no longer dominate the energetic content.
In the antisymmetric vibration regime, a very similar modal structure is observed. The first and second modes remain antisymmetric, reinforcing the dominance of antisymmetric flow–structure interaction. The third and fourth modes retain symmetric structures, indicating that symmetric fluctuation patterns persist in parts of the wake despite the dominant antisymmetric dynamics.
This progression shows that mode shapes evolve distinctly with the speed and regime, directly reflecting changes in the flow and synchronization with the plate deformation. The coherent flow structures consistently follow the plate’s deformation, with symmetry or antisymmetry observed in both the flow and the plate vibration. In the transition regime, visual observations of the plate during acquisition clearly revealed the simultaneous presence of both symmetric and antisymmetric vibrations. This coexistence of competing structural dynamics confirms the transitional nature of the flow–structure interaction. Shifts between symmetric and antisymmetric dominance in the POD modes serve as indicators of evolving vibration regimes and associated energy redistribution, highlighting the need to also examine how these modes evolve with flow speed and how their energetic contributions vary.
In this regard, the symmetrical and antisymmetrical type wake signatures identified in Figure \ref{fig:mode shapes} can be followed across the regimes through their shifting energy levels. For instance, the symmetrical mode dominates in the symmetrical vibration regime but reappears with secondary energy in the antisymmetrical regime, maintaining the same coherent spatial structure. Conversely, the antisymmetrical mode emerges with secondary energy during the symmetrical regime before becoming dominant in the antisymmetrical regime. This continuity of spatial patterns, coupled with their changing energetic weight, emphasises the robustness of the underlying coherent structures and illustrates how energy redistribution between symmetrical structure and antisymmetrical structure reflects the transitions in wake–plate interaction.

\begin{figure}
\centering
\normalfont
\def\svgwidth{1\textwidth}
\input{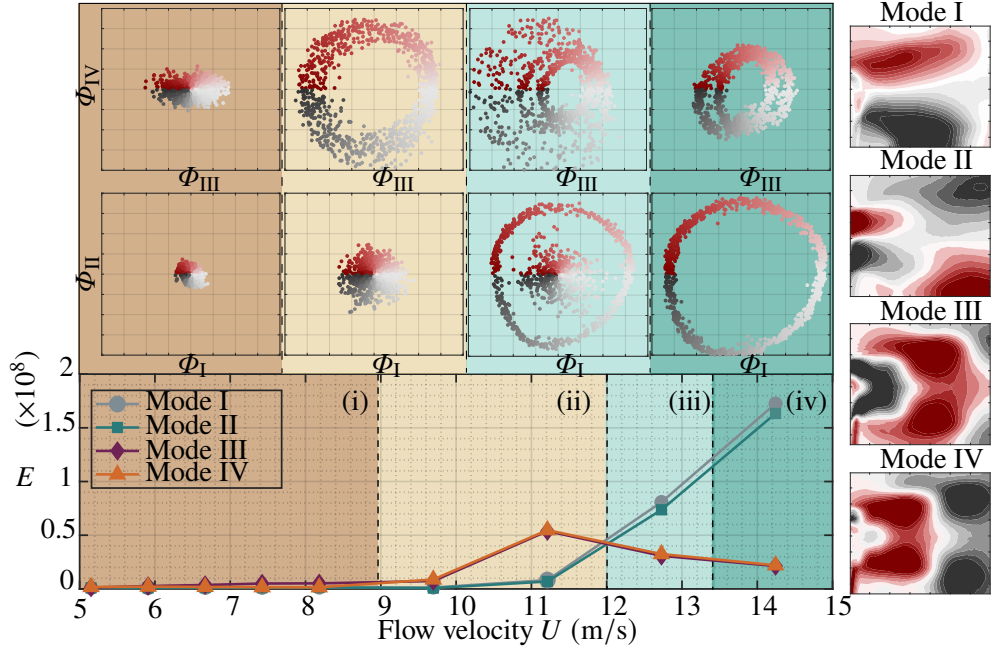}
\caption{Evolution of modal dynamics and energy distribution across flow regimes. 
\textbf{Top:} Phase portraits constructed from the temporal coefficients (modal amplitudes) of the first four POD modes of the streamwise velocity field. Each subplot shows the projection in the $(\Phi_I, \Phi_{II})$ or $(\Phi_{III}, \Phi_{IV})$ plane for four flow regimes: (a) static reconfiguration ($U$ = 8.2~\si{\meter\per\second}), (b) symmetric vibration ($U$ = 11.2~\si{\meter\per\second}), (c) transition regime ($U$ = 12.7~\si{\meter\per\second}), and (d) anti-symmetric vibration ($U$ = 14.2~\si{\meter\per\second}). \textbf{Bottom:} Energy contribution of the first four POD modes as a function of flow speed. The coloured background identifies the dominant flow regime: (i) static reconfiguration (light brown), (ii) symmetric vibration (beige), (iii) transition regime (pale aqua), and (iv) anti-symmetric vibration (green-blue). }
\label{fig:global modes}
\end{figure}

To systematically compare the flow dynamics across regimes, we introduce global combined modes. These are obtained by applying the POD on the full dataset encompassing all flow velocities, rather than analysing each case independently. In this way, a single set of spatial modes spans the entire parameter space, allowing us to track how the same coherent structure redistributes its energy contribution as the flow transitions from static to vibrating regimes.
Figure \ref{fig:global modes} presents the energy contribution of the first four global combined modes as a function of flow velocity, alongside their phase portraits constructed from the corresponding temporal coefficients.

To evaluate the global contribution of modes across different experimental velocities, the individual data matrices $L_{p=1}, L_{p=2}, L_{p=3},...,L_{p=k}$ ($k$ being the total number of flow speeds considered) are concatenated into a single matrix, denoted as $L_{combined}$ . Singular Value Decomposition (SVD) is applied to decompose $L_{combined}$ into three matrices: $\Psi_f$, $\Sigma_f$, and $\Phi_f$. 
The modes extracted from SVD of the concatenated data matrix, which integrates all velocity cases, are referred to as global combined modes. These modes are subsequently projected onto each individual velocity data matrix $L_p$ to evaluate their contribution at each flow condition:
\begin{equation}
(Z_p)_{k j}
=
\sum_{i=1}^{2N}
\Psi_{i k}\,(L_p)_{i j},
\qquad
k = 1,\ldots,r\;\;,\;\; j = 1,\ldots,m,
\end{equation}
where $r$ is the number of global combined modes considered, and  $m$ is the number of snapshots at speed $p$. The energies for each mode are given without normalization, so differences between speeds are retained:
\begin{equation}
    (E_{p})_k = \sum_{j=1}^{m} (Z_{p})_{kj}^2 
    \qquad k=1,\ldots,r .
\end{equation}

Using this framework, the analysis of the global combined modes spanning the full range of experimental velocities is displayed in Figure \ref{fig:global modes}, which offers an in-depth perspective on the dynamic evolution of the system, including the contributions of modal energy and the phase dynamics characterising each regime.
The lower part of Figure \ref{fig:global modes} shows how the energy contributions of the global combined modes evolve with increasing flow velocity. Four distinct regimes emerge from this analysis. At low velocities, corresponding to the static reconfiguration regime, all modes exhibit negligible energy contributions, indicating that the structure remains essentially static with no significant dynamic excitation. As velocity increases, the system transitions into the symmetric vibration regime, where modes III and IV, associated with symmetric structural deformations, become dominant. Their energy contributions rise sharply, indicating the onset of symmetric oscillations. In contrast, modes I and II, related to antisymmetric structures, exhibit a slight increase compared to the previous regime but remain substantially lower than modes III and IV, suggesting limited anti-symmetric activity at this stage. Continuing to increase the velocity leads to the transition regime, characterised by a progressive decay in the contributions of modes III and IV and a simultaneous rise in modes I and II. This regime reflects the coexistence and competition between symmetric and anti-symmetric dynamics as the system approaches a bifurcation point. Finally, at higher velocities, the system reaches the antisymmetric vibration regime, where modes I and II clearly dominate, and the energy contributions of modes III and IV become negligible, indicating a shift to fully antisymmetric structural behaviour.

Complementing the energy analysis, the higher panels of Figure \ref{fig:global modes} present phase portraits that capture the dynamical relationships between modal amplitudes. For each flow regime, these portraits illustrate the trajectories in two modal planes: (mode I, mode II) for antisymmetric structures and (mode III, mode IV) for symmetric structures. In the static reconfiguration regime, both phase portraits exhibit disorganized clouds of points, indicative of the absence of coherent oscillatory dynamics. In contrast, during the symmetric vibration regime, the phase portrait in the (mode III, mode IV) plane forms a well-defined circular trajectory, confirming the presence of stable symmetric limit-cycle oscillations, while the (mode I, mode II) plane remains a scattered point cloud. As the flow enters the transition regime, the $\Phi_{I} - \Phi_{II}$ portrait evolves into a configuration where a central cluster of points is surrounded by a circular trajectory, reflecting the emergence of anti-symmetric oscillations superimposed on residual symmetric behaviour. Meanwhile, the phase plane between $\Phi_{III}$ and $\Phi_{IV}$ shows a broad circular pattern with multiple overlapping loops, indicating periodic oscillations between the two modes, but with amplitude variations over time.  This spread of trajectories suggests modulations, preventing the oscillations from being confined to a single amplitude. The system remains within a stable oscillatory regime, but with fluctuating amplitudes. Finally, in the antisymmetric vibration regime, the $\Phi_{I} - \Phi_{II}$ phase portrait presents a clear and coherent circular trajectory, marking the dominance of antisymmetric oscillations. The $\Phi_{III}$-$\Phi_{IV}$ phase plane reveals a deformed figure-eight pattern. This shape suggests that modes III and IV are coupled and oscillate with varying relative amplitudes, possibly influenced by interactions with the dominant antisymmetric mode. Although the trajectory remains bounded, its spread reflects temporal variations in oscillation amplitude.
Together, these analyses of energy contributions and phase dynamics provide a comprehensive understanding of how the system transitions across different flow-structure interaction regimes, highlighting the interplay between symmetric and antisymmetric modes as the flow velocity varies.

\subsection{Periodic vorticity and wake pattern reconstruction}
\begin{figure}
\centering
\normalfont
\def\svgwidth{1.0\textwidth}
\input{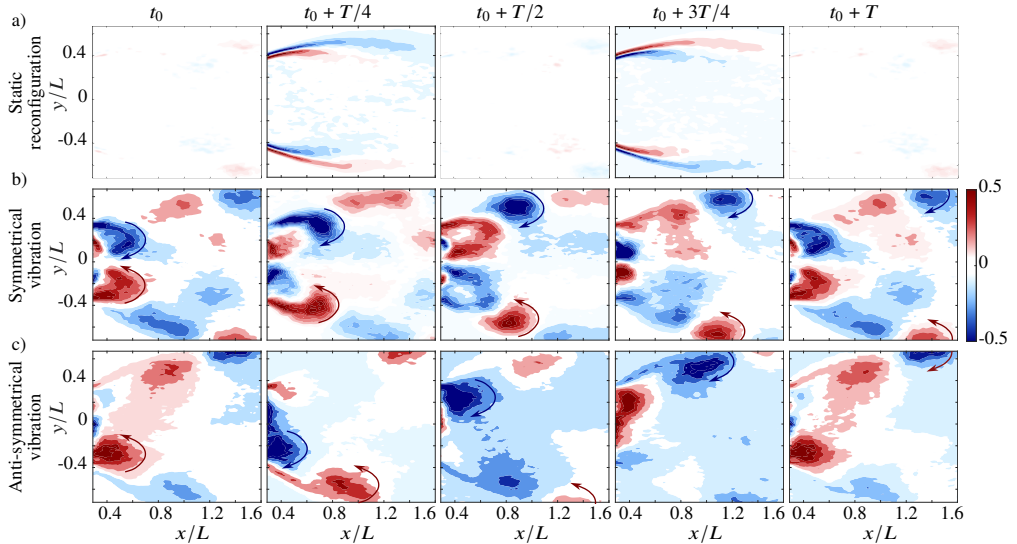}
\caption{Spanwise vorticity fields $\omega_z$ at five successive time instants ($t_0$, $t_0+T/4$, $t_0+T/2$, $t_0+3T/4$, $t_0+T$) for three flow regimes: a) static reconfiguration; b) symmetric vibration; and c) anti-symmetric vibration. Here, $t_0$ marks the initial instant chosen for visualization, and $T$ denotes the oscillation period. }
\label{fig:vorticity}
\end{figure}

Phase portraits and energy evolution establish the modal organisation across regimes; to resolve the spatial dynamics, we turn to the spanwise vorticity. The vorticity field $\omega_z(x,y,t)$ is computed from reconstructed flow field of Eq.~\ref{eq: reconstructed M} as the curl of the velocity:
\begin{equation}
\omega_z(x,y,t) = \frac{\partial v}{\partial x} - \frac{\partial u}{\partial y}.
\end{equation}
To this end, Figure~\ref{fig:vorticity} presents the spanwise vorticity fields at five successive time instants over one oscillation period for three representative regimes: static reconfiguration, symmetric vibration, and antisymmetric vibration. These visualisations reveal how the wake topology evolves with increasing flow speed and structural vibration, offering a clearer picture of the vortex shedding patterns associated with each regime.

In the static reconfiguration regime, the vorticity field remains largely unperturbed, displaying a stable mean wake with minimal unsteady features. This reflects the absence of significant dynamic fluid-structure interactions: the plate maintains a steady deflection. Consistently, the phase-averaged vorticity fields show their strongest signal at $t_0+T/4$ and $t_0+3T/4$, where opposite-signed vorticity develops along the separated shear layers, while the fields at $t_0$, $t_0+T/2$, and $t_0+T$ remain comparatively weak. Mode I in Figure \ref{fig:mode shapes} is symmetric and captures the dominant large-scale wake fluctuation, with structures concentrated along the separated shear layers on either side of the plate. Overall, this spatial organization is consistent with shear-layer-dominated wake dynamics, without a clear signature of a dominant, phase-locked vortex-shedding mode in the present near-wake measurements \citep{wiliamson1996vortex}. Because the present field of view is limited to the near wake, the development of a fully formed von Kármán vortex street further downstream cannot be excluded.
In the symmetric vibration regime, the interaction between the oscillating plate and the incoming flow leads to the periodic shedding of counter-rotating vortex pairs from either side of the structure. At time $t_0$, a positive vortex (red) is released from one side of the plate, while a negative vortex (blue) is simultaneously shed from the opposite side, forming a symmetric pair. As they convect downstream, the vortices gradually move away from the wake centreline, a behaviour likely influenced by the transverse vibrations of the flexible plate during shedding.
At time $t_0+T/4$, a new pair of vortices is shed, with opposite circulation compared to the previous pair. This alternating sequence continues over the cycle, resulting in a regular pattern of symmetric vortex shedding. The periodic and mirror-like nature of the shedding is consistent with the symmetric deformation of the plate observed in this regime, as also reflected in the dominant POD mode structures.

The anti-symmetric vibration regime reveals a markedly different fluid-structure interaction. Here, the plate undergoes anti-symmetric deformation, inducing a sequential vortex shedding mechanism. A pair of vortices (positive followed by negative) is first released from one side of the plate as it deflects toward that side. Shortly after, as the plate deflects to the opposite side, another pair of vortices with opposite signs is shed from the other side. These vortices convect downstream along the wake edges, creating an alternating shedding pattern synchronized with the anti-symmetric motion of the structure. The modal analysis corroborates this behaviour, with anti-symmetric modes dominating the energy contribution in this regime. The alternating vortex shedding observed in the wake induces fluctuating forces on the structure, a mechanism commonly associated with vortex-induced vibrations in flexible bodies \citep{williamson2004vortex}.

\begin{figure}
\centering
\normalfont
\def\svgwidth{0.9\textwidth}
\begingroup%
  \makeatletter%
  \providecommand\color[2][]{%
    \errmessage{(Inkscape) Color is used for the text in Inkscape, but the package 'color.sty' is not loaded}%
    \renewcommand\color[2][]{}%
  }%
  \providecommand\transparent[1]{%
    \errmessage{(Inkscape) Transparency is used (non-zero) for the text in Inkscape, but the package 'transparent.sty' is not loaded}%
    \renewcommand\transparent[1]{}%
  }%
  \providecommand\rotatebox[2]{#2}%
  \newcommand*\fsize{\dimexpr\f@size pt\relax}%
  \newcommand*\lineheight[1]{\fontsize{\fsize}{#1\fsize}\selectfont}%
  \ifx\svgwidth\undefined%
    \setlength{\unitlength}{898.58691598bp}%
    \ifx\svgscale\undefined%
      \relax%
    \else%
      \setlength{\unitlength}{\unitlength * \real{\svgscale}}%
    \fi%
  \else%
    \setlength{\unitlength}{\svgwidth}%
  \fi%
  \global\let\svgwidth\undefined%
  \global\let\svgscale\undefined%
  \makeatother%
  \begin{picture}(1,0.40206679)%
    \lineheight{1}%
    \setlength\tabcolsep{0pt}%
    \put(0,0){\includegraphics[width=\unitlength,page=1]{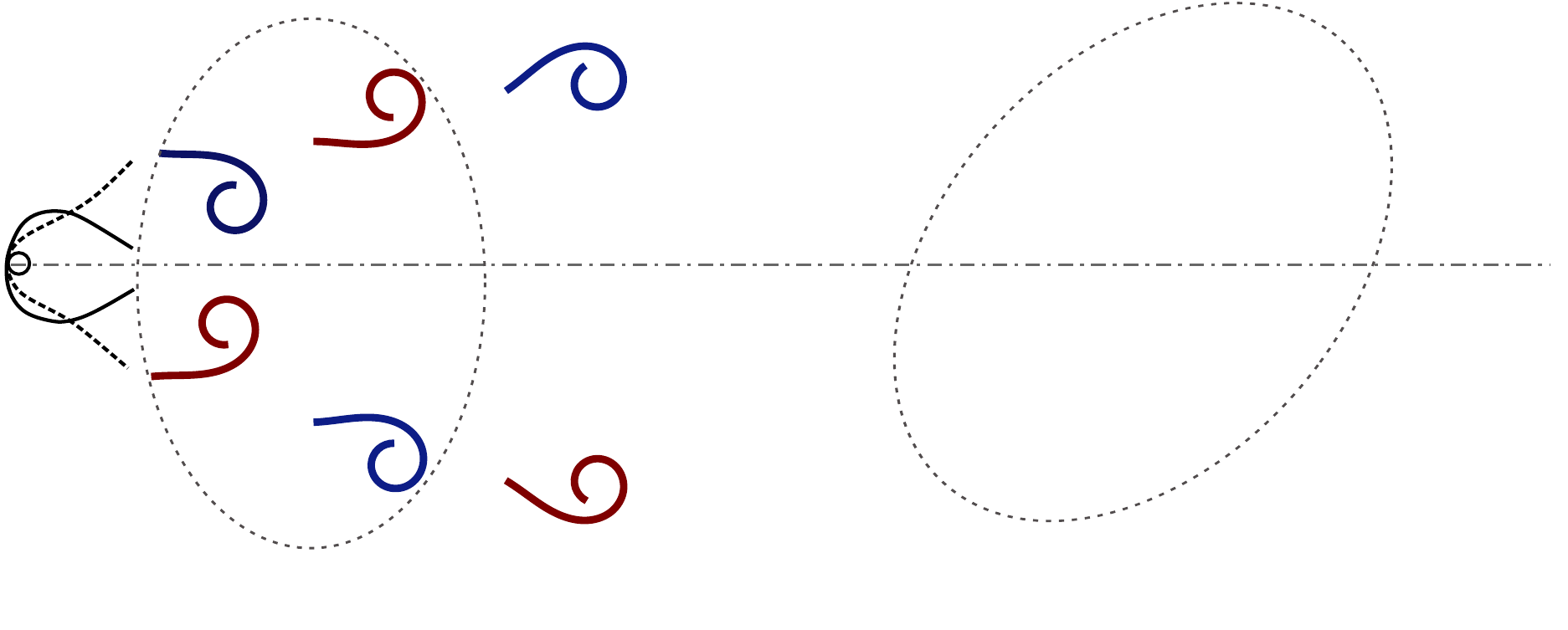}}%
    \put(0.76422854,0.22){\makebox(0,0)[t]{\lineheight{1.25}\smash{\begin{tabular}[t]{c}2P\end{tabular}}}}%
    \put(0.25,0.22){\makebox(0,0)[t]{\lineheight{1.25}\smash{\begin{tabular}[t]{c}S-2S\end{tabular}}}}%
    \put(0,0){\includegraphics[width=\unitlength,page=2]{Figures/figure08.pdf}}%
    \put(0.02067535,0.37241734){\makebox(0,0)[t]{\lineheight{1.25}\smash{\begin{tabular}[t]{c}a)\end{tabular}}}}%
    \put(0.49678597,0.37241681){\makebox(0,0)[t]{\lineheight{1.25}\smash{\begin{tabular}[t]{c}b)\end{tabular}}}}%
    \put(0,0){\includegraphics[width=\unitlength,page=3]{Figures/figure08.pdf}}%
  \end{picture}%
\endgroup%

\caption{Schematics of the vortex shedding patterns found for: a) symmetrical vibration regime; and b) anti-symmetrical vibration regime. \textbf{P}: vortex pair, \textbf{S}: Single vortex. }
\label{fig: sketch wake}
\end{figure}

To facilitate the interpretation of the complex wake patterns observed in the vorticity fields, we introduce in Figure \ref{fig: sketch wake}  a set of schematic representations of the dominant vortex configurations for each vibration regime. 
These sketches provide a simplified visualisation of the main vortex shedding mechanisms including symmetric and antisymmetric release and serve as a conceptual summary of the wake dynamics associated with each structural mode.
To interpret the symmetric shedding mechanism, complementary analogies can be drawn.
First, a comparison can be made with the classical 2S vortex shedding mode observed in oscillating cylinders in the cross-stream direction, where two single vortices of opposite sign are shed alternately from each side of the body during each oscillation cycle \citep{williamson1988vortex}. In the present configuration, this mechanism manifests as a 2S-type pattern developing on each side of the vibrating plate.

A related analogy can be drawn with the streamwise response regime observed in cylinders undergoing vortex-induced vibrations, where vortex pairs are generated simultaneously on both sides of the body due to the periodic inward–outward motion of the separating shear layers, commonly referred to as wake breathing. This regime corresponds to the S-I vortex pattern identified in streamwise VIV studies, in which the shear layers roll up synchronously at the forcing frequency \citep{ongoren1988flow, cagney2014streamwise}. 
In contrast to the classical S-I mode, which produces pairs of identical sign in the near wake, the present plate configuration exhibits an alternation of vortex sign between successive pairs. Similar behaviour has been reported in cylinders subjected to externally imposed streamwise perturbations, where strong entrainment of outer fluid between successive vortex pairs significantly alters the local vorticity distribution. Phase-averaged analyses in these studies reveal pronounced deviations in the fluctuating vorticity field, including local changes in sign associated with the entrainment process, although this behaviour is observed only under specific forcing conditions \citep{konstantinidis2007symmetric}. 
Based on these observations, the wake configuration observed here can be interpreted as a symmetric arrangement of two parallel 2S-type shedding patterns developing on each side of the plate. In the following, this configuration is referred to as a S-2S mode, denoting a symmetric wake composed of two adjacent 2S vortex streets. The symmetric shedding is also accompanied by a periodic contraction and expansion of the wake, reminiscent of the wake-breathing dynamics reported in streamwise VIV studies. Additional evidence of this wake breathing behaviour in the plate wake is provided in Appendix \ref{appA}.

In addition, a biologically inspired analogy can be established with the propulsion mechanism of pulsating jellyfish. During each contraction cycle, the flow organises into axisymmetric vortex rings shed from the bell \citep{hoover2017quantifying}. When viewed in planar sections, these vortex arrangement resembles the symmetric 2S shedding observed in the plate’s wake during the symmetrical vibration. Such an analogy highlights how similar flow structures can emerge from distinct physical mechanisms, one biologically driven and the other mechanically induced, yet governed by comparable symmetry and unsteady forcing.

In the case of antisymmetric vibrations, the wake structure is analogous to the classic 2P shedding mode observed in oscillating cylinders, where pairs of vortices are formed and convected laterally away from the wake centreline \citep{williamson1988vortex}.

\subsection{Impulse-based wake mean drag contribution}
\begin{figure}
\centering
\normalfont
\def\svgwidth{0.9\textwidth}
\input{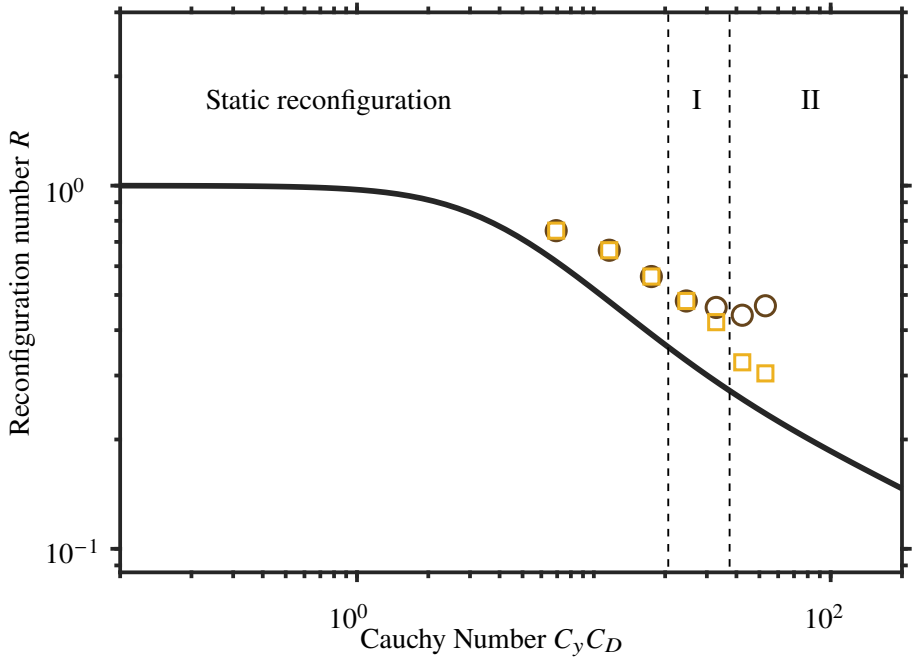}
\caption{Mean reconfiguration number $R$ versus Cauchy number $C_yC_D$, with and without impulse-based correction. \scalebox{1.5}{$\circ$}: force balance measurements; $\square$: values corrected by subtracting the wake mean drag estimated from the impulse theory. The black curve ($\textcolor[rgb]{0,0,0}{\rule{0.5cm}{0.5mm}}$) represents  the model of \cite{gosselin2010drag}. I: symmetric vibration, II: antisymmetric vibration.
 }
\label{fig: induc}
\end{figure}

The wake visualizations presented in the previous section highlight the profound changes in vortex organization across regimes, but they also raise the question of how such unsteady dynamics translate into the measured aerodynamic forces. The drag is not only determined by the plate’s reconfigured shape, but also by the way its flapping motion interacts with the vortices in the wake, which can add extra drag. This issue became apparent in the measured time-averaged drag of a similar plate system by \cite{boukor2024flutter} and reproduced in Figure \ref{fig: induc} with circle symbols. The reconfiguration number is defined as the ratio between the drag of the flexible structure and the drag of an identical rigid plate  experiencing under the same flow conditions. The rigid drag is obtained from wind-tunnel measurements on rigid models by determining the drag coefficient for plates of various dimensions:
\begin{equation}
    \label{eq:reconfigurationnumber}
R \;=\; \frac{F_x}{\tfrac{1}{2}\rho_f U^2 C_D A},
\end{equation}
where $F_x$ is the measured mean drag with a force balance, $C_D$ the drag coefficient of the rigid plate, and $A$ its undeformed frontal area. The results showed that the reconfiguration number obeys a constant scaling with the Cauchy number of $R\propto C_y^{-1/3} $ for static and symmetric regimes, but the antisymmetric regime showed a marked deviation in the time-averaged drag measured by the force balance  (Figure \ref{fig: induc}, circle symbols). This deviation suggests that the anti-symmetric flapping gives rise to a significant increase in time-averaged drag.

This discrepancy suggests that another mechanism obeying a different scaling is generating drag during antisymmetric flapping. Our working hypothesis is that, during antisymmetric flapping, the periodically shed vortices generate wake circulation, which creates an additional drag component with a non-zero mean. 
If this contribution is removed from the balance measurements, the corrected drag should collapse back onto the same scaling as the static and symmetric cases. Testing this hypothesis requires a complementary approach based on impulse force theory, where wake circulation and body motion can be explicitly related to the mean drag. In the present work, the impulse-based framework is used primarily as a diagnostic tool to isolate the wake-vorticity contribution in the unsteady regimes, rather than as a strict estimate of the absolute mean drag.

The drag force was estimated following \cite{wu1981theory}’s impulse formulation, which relates the aerodynamic force to the first moment of vorticity in the wake. Impulse-based formulations belong to a broader class of exact force expressions derived from velocity-field measurements, whose equivalence and respective advantages have been discussed extensively in the literature (see, e.g., \cite{rival2017load}; \cite{limacher2019added}). Among these approaches, vorticity-based formulations are particularly attractive for experimental applications, as they avoid pressure reconstruction and allow a direct interpretation of aerodynamic forces in terms of wake dynamics. It has further been noted by \cite{graham2017impulse} that, when impulse-based force estimates are found to be in close agreement with direct force measurements, this agreement implies a predominantly two-dimensional unsteady flow, thereby providing an a posteriori justification for neglecting the three-dimensional induced-drag contribution.
For a two-dimensional configuration of span $w$, the streamwise force per unit span is
\begin{equation}
F_D(t) \;=\; -\,\rho_f\,\frac{d}{dt}\iint_{\mathcal{A}} y'\,\omega_z(x,y,t)\,dx\,dy,
\label{eq: wu's equation}
\end{equation}
where  $\omega_z$ is the spanwise vorticity, $\mathcal{A}$ the measurement window in the $(x,y)$ plane, and $y' \;=\; y - y_0(t)$  is the vertical position of the plate, which we estimated from the motion tracking and the centroid motion model of Eq. (\ref{eq:centroid}). 
The velocity fields $(u,v)$ were obtained from the PIV reconstructions and interpolated on a regular grid. For each pixel, the temporal signal of $\omega_z$ over one oscillation was fitted by
\begin{equation}
\omega_z(x,y,t) \;\approx\; \omega_0(x,y)\sin\!\big(2\pi ft) + \bar{\omega}_z(x,y).
\end{equation}
 This fitting isolates the phase-locked amplitude $\omega_0(x,y)$ and the mean vorticity $\bar{\omega}_z(x,y)$ of the vorticity oscillation. 
 Inserting these forms into Wu’s expression (\ref{eq: wu's equation}) and expanding the derivative separates the drag into three contributions: an oscillatory term proportional to $\sin^2(\theta)$, a term proportional to $\cos(\theta)$, and a constant contribution. The constant term, denoted $D_c$ in the implementation, is
\begin{equation}
{\, D_c \;=\; 2\pi\,\rho_f\,w\,f\,\iint_{\mathcal{A}}Y_0\,\omega_0 \,dx\,dy\,},
\end{equation}
with $\rho_f$ the air density, $f$ the flapping frequency and $w$ the plate's width. Among the three contributions, the first two are oscillatory and integrate to zero over one period, leaving only the constant third term modifying the mean drag. This correction vanishes in the static and symmetric regimes, where the centroid motion $y_0$ is negligible, but becomes significant in the antisymmetric regime, where $y_0 \neq 0$. A small exception occurs at $U$=11.2~ \si{\meter\per\second} in the nominally symmetric regime: as shown in the mode-contribution analysis (Figure ~\ref{fig:global modes}), the antisymmetric mode gains a slight increase in energy compared to the previous point. This weak antisymmetric component is consistent with the small but non-zero $y_0$ obtained from image tracking, and it produces a minor correction despite the overall symmetry of the wake.

Figure~ \ref{fig: induc} presents the evolution of the reconfiguration number $R$ as a function of the Cauchy number $C_yC_D$. Here the Cauchy number is scaled with the drag coefficient, following the model proposed by \cite{gosselin2010drag}, and the theoretical curve representing the model prediction for $R$ is also shown in the figure.
Two sets of values are reported. The first set corresponds to the raw balance data, i.e.\ the mean drag force measured directly with the load cell and made dimensionless with Eq. (\ref{eq:reconfigurationnumber}). The second set includes a correction for the induced mean contribution identified in the impulse analysis, obtained by subtracting the constant term $D_c$ from the balance drag. As hypothesized, the static and symmetric regimes exhibit the same trend as the theoretical curve. However, a systematic gap is observed due to discrepancies between the wind-tunnel measurements and the numerical reference used in the model. In the symmetric regime, the same trend is observed overall, although a slight correction is required at certain velocities where a small antisymmetric contribution is present. Once corrected, these points fall into the same trend as in the static case. By contrast, the antisymmetric regime shows a marked deviation when using the raw balance data, but the corrected values recover the same trend as the static and symmetric regimes, supporting the interpretation that the excess drag originates from oscillation-induced circulation.

Overall, this comparison shows that once the impulse-based wake-induced mean drag contribution is accounted for, the  antisymmetric regime recovers the same reconfiguration trend as the static and symmetric cases, indicating that the scaling framework for the mean drag can be extended to all regimes in the present experiments.

\section{Conclusion}

In this study, we investigated the wake dynamics of flexible structures that progressively transition from cross-flow bluff bodies to axial-flow vibratory instability. While wake patterns are well documented for rigid bluff bodies in cross-flow, and stability analyses exist for fluttering axial-flow flags, direct wake measurements for hybrid structures that progressively reconfigure and then start vibrating remain scarce. We described how the near wake develops from static reconfiguration into dynamic vibration-driven regimes, and how these changes directly translate to the measured aerodynamic forces. By applying a modal decomposition and phase-ordered reconstruction approach to independent PIV snapshots, we successfully extracted coherent wake structures and regime-dependent signatures despite the lack of initial temporal resolution.

The wake topology was shown to depend directly on the vibration regime of the plate. Symmetric oscillations produced a symmetric vortex release, corresponding to a wake structure composed of two parallel 2S-type shedding patterns on either side of the plate, herein referred to as an S-2S mode. This configuration also bears similarity to the paired vortex structures observed in pulsating jellyfish. In contrast, antisymmetric oscillations led to alternating vortex pairs, analogous to the 2P shedding mode in cylinders. These analogies emphasize the close link between structural symmetry and wake organisation.

Finally, by complementing force balance measurements with an impulse-based force estimate, we identified a non-zero mean drag associated with centroid oscillations and wake circulation in the antisymmetric regime. More generally, the vibration regimes were found to generate pronounced dynamic loads through the fluctuating drag coupled to the motion, whereas only the antisymmetric regime exhibited an additional non-zero mean drag contribution. Subtracting this contribution restored the collapse of the drag data onto the reconfiguration scaling that holds for static and symmetric cases. This correction provides a mechanistic explanation for the excess drag observed in the antisymmetric regime and further supports the relevance of impulse theory for connecting wake dynamics to aerodynamic loads.

While the present work clarifies how wake organisation and aerodynamic loading evolve across vibration regimes in a reconfiguring flexible plate, this foundation opens several new avenues for further investigation. Future work will deploy the impulse-based force approach to other canonical flexible systems, including flags, to fundamentally relate regime-dependent wake dynamics to mean and fluctuating loads. Furthermore, applying time-resolved PIV will recover instantaneous spectral content and phase relationships beyond snapshot reconstructions, while volumetric or mutually perpendicular two-plane PIV measurements will resolve critical spanwise three-dimensional effects. Broadening the parameter space via numerical simulations will test the robustness of the mechanisms identified here, and complementary experiments at larger scales are planned to generalise these findings beyond the current geometric constraints.


\section*{Funding}
The financial support of the Natural Sciences and Engineering Research Council of Canada, the Canada Foundation for Innovation, and the Canada Research Chair Program is acknowledged.

\begin{appen}
\section{}\label{appA}
The evolution of the wake width, shown in Figure \ref{fig: appendix}, is examined in order to illustrate the periodic contraction and expansion of the wake during the symmetric shedding regime, commonly referred to in the literature as wake breathing. The wake boundary is identified using the velocity-deficit contour $u/U = 0.8$. For each streamwise position, the wake width is defined as the vertical distance between the upper and lower intersections of this contour. The resulting profiles of the normalized wake width $W_{\text{wake}}/L$ are shown for several snapshots taken around the half-cycle of the oscillation, including the phase $j=T/2$. The curves show that the wake progressively narrows downstream before expanding again further downstream. This behaviour reflects the inward–outward motion of the separating shear layers during the oscillation cycle. The instantaneous velocity field at $j=T/2$, shown in the figure together with the contour used to identify the wake boundary, corresponds to the phase where the wake contraction is most pronounced.

\begin{figure}
\centering
\normalfont
\def\svgwidth{0.9\textwidth}
\input{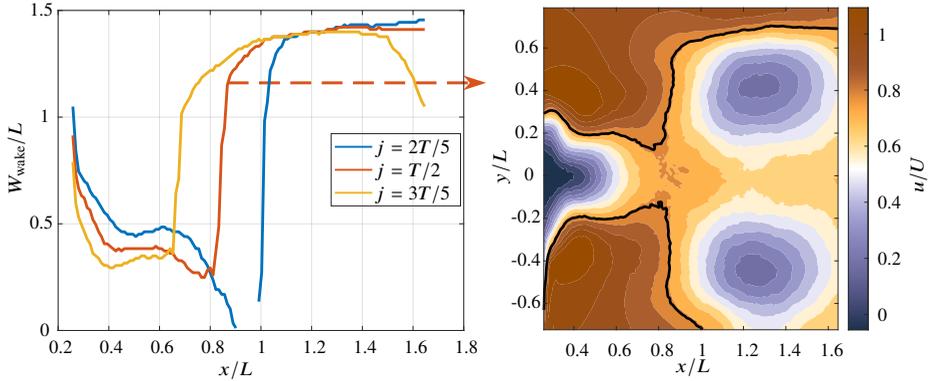}
\caption{ Wake breathing during the symmetric shedding regime. 
\textbf{Left:} Streamwise evolution of the normalized wake width $W_{\text{wake}}/L$ for snapshots around the half-cycle of oscillation. The wake width is defined as the vertical extent of the velocity-deficit region identified by the contour $u/U = 0.8$ (black contour). 
\textbf{Right:} Instantaneous streamwise velocity field at the phase $j=T/2$, where the wake contraction is most pronounced. $T$ denotes the oscillation period. }
\label{fig: appendix}
\end{figure}

\section{Randomized Dependence Coefficient (RDC)}\label{sec:appendix2}
Before introducing the reconstructed data in Section 3, a correlation analysis based on Randomized Dependence Coefficient (RDC) was performed on the POD coefficients extracted from the cleaned snapshots obtained after RPCA, corresponding to the matrix $\boldsymbol{L}$ in Eq.~\ref{eq: matrix L}, and prior to any sorting or fitting procedure. 

The purpose of the RDC analysis is to verify the underlying assumption that enables the phase-based sorting procedure. The sorting strategy relies on the existence of statistical dependence between POD modes: if the modes are dynamically correlated, their joint distributions should exhibit coherent Lissajous-type structures, indicating that they share a common underlying phase evolution. In such a case, defining a phase angle from the first two modes is expected to meaningfully organise the snapshots not only for these modes but also for the higher-order ones.

Conversely, if the RDC analysis reveals only unstructured point clouds, indicating weak or absent intermodal dependence, the phase-based sorting would not be expected to succeed. In that situation, organising snapshots according to the phase of modes 1 and 2 would not produce coherent behaviour in the remaining modes. The reconstruction procedure itself does not create additional information; it merely rearranges existing snapshots. Therefore, the effectiveness of the sorting step critically depends on the presence of intrinsic modal coherence in the cleaned dataset.

Originally introduced by \cite{lopez2013randomized}, the RDC is a nonlinear dependence measure that can capture both linear and nonlinear relationships, while remaining invariant to monotonic transformations and amplitude scaling. Its relevance for the analysis of modal interactions in reduced-order models has been further emphasised by \cite{callaham2022role}.
The RDC relies on a combination of rank-based transformations and randomized nonlinear projections. Given two variables, their samples are first mapped onto their empirical copula representation to remove the influence of marginal distributions. Nonlinear features are then generated using sinusoidal projections with randomly sampled frequencies and phases. Canonical correlation analysis is subsequently applied to these features, and the maximum canonical correlation defines the RDC value.
The resulting coefficient ranges between 0 and 1, with values close to zero indicating weak statistical dependence and values close to unity indicating strong dependence, irrespective of whether the underlying relationship is linear or nonlinear. In the present study, the RDC is used as a diagnostic tool to assess statistical dependencies between pairs of POD temporal coefficients. For clarity, RDC values are presented alongside the corresponding phase portraits. Figures \ref{fig: RDC symmetrical regime}, \ref{fig: RDC transition regime}, and \ref{fig: RDC antisymmetrical regime} show the results obtained for the symmetrical, transition, and antisymmetrical regimes, respectively.

These results show that significant statistical dependence exists between the dominant modes, and that higher-order modes remain correlated with modes 1 and 2. The structured RDC phase planes provide evidence that coherent intermodal dynamics are already present, thereby justifying the use of a phase defined from the first two modes to organise the entire modal set.
An exception is observed for antisymmetric structures when the flow is dominated by a symmetric instability. In this specific regime, antisymmetric modes appear weakly correlated with modes 1 and 2, highlighting a limitation of the approach.
Overall, the observed modal correlations demonstrate that the leading POD modes are dynamically linked, which justifies the use of a phase angle constructed from their temporal coefficients for snapshot ordering.

\begin{figure}
\centering
\normalfont
\def\svgwidth{1\textwidth}
\input{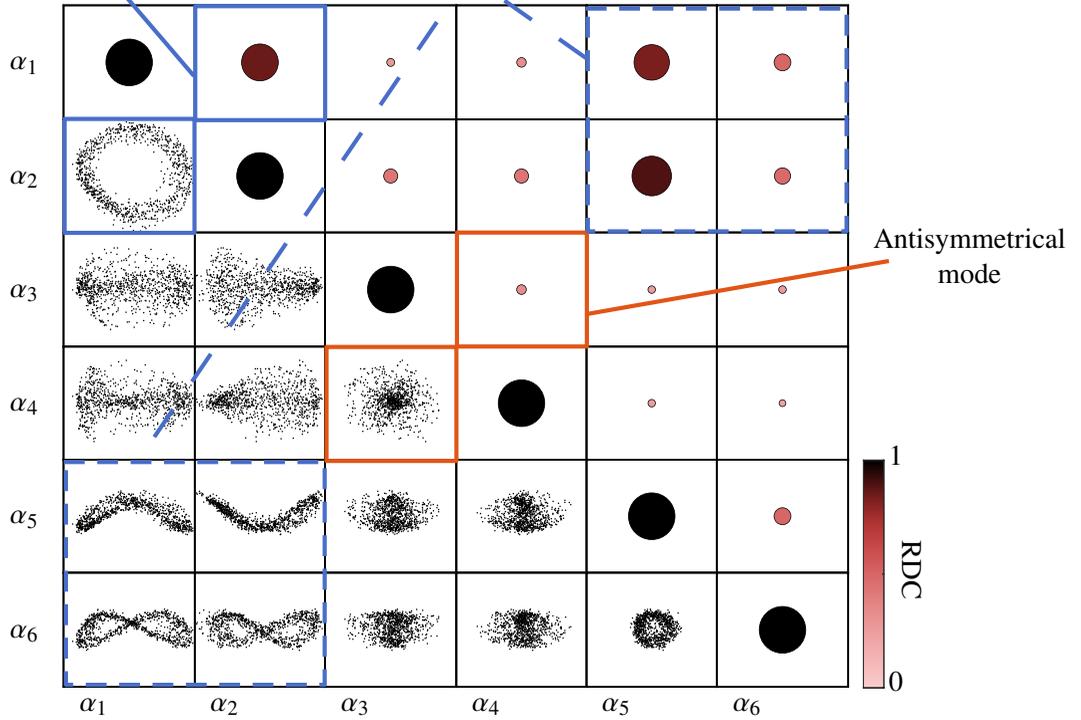}
\caption{Correlation analysis of POD temporal coefficients using the Randomized Dependence Coefficient (upper triangle) and phase portraits (lower triangle). Symmetric modes are shown in blue, antisymmetric modes in orange, and the corresponding harmonics are indicated by dashed outlines. These results correspond to the symmetrical regime at $U=11.2m/s$}
\label{fig: RDC symmetrical regime}
\end{figure}

\begin{figure}
\centering
\normalfont
\def\svgwidth{1\textwidth}
\input{Figures/figure12t.tex}
\caption{Correlation analysis of POD temporal coefficients using the Randomized Dependence Coefficient (upper triangle) and phase portraits (lower triangle). Symmetric modes are shown in blue, antisymmetric modes in orange, and the corresponding harmonics are indicated by dashed outlines. These results correspond to the transition regime at $U=12.7m/s$}
\label{fig: RDC transition regime}
\end{figure}

\begin{figure}
\centering
\normalfont
\def\svgwidth{1\textwidth}
\input{Figures/figure13t.tex}
\caption{Correlation analysis of POD temporal coefficients using the Randomized Dependence Coefficient (upper triangle) and phase portraits (lower triangle). Symmetric modes are shown in blue, antisymmetric modes in orange, and the corresponding harmonics are indicated by dashed outlines. These results correspond to the antisymmetrical regime at $U=14.2m/s$}
\label{fig: RDC antisymmetrical regime}
\end{figure}
\end{appen}\clearpage

\bibliographystyle{jfm}
\bibliography{jfm}

\end{document}